# Bicycle Longitudinal Motion Modeling

**Karim Fadhloun, Ph.D.**
Postdoctoral Associate, Center for Sustainable Mobility, Virginia Tech Transportation Institute
Center for Sustainable Mobility
Virginia Tech Transportation Institute, Blacksburg, Virginia 24061
Email: karim198@vt.edu

**Hesham Rakha, Ph.D., P.Eng. (Corresponding author)**
ORCiD: https://orcid.org/0000-0002-5845-2929
Samuel Reynolds Pritchard Professor of Engineering and Director
Center for Sustainable Mobility
Virginia Tech Transportation Institute, Blacksburg, Virginia 24061
Email: hrakha@vt.edu

**Archak Mittal, Ph.D.**
ORCiD: https://orcid.org/0000-0001-6186-4513
Research Scientist
Ford Motor Company
20000 Rotunda Drive
Dearborn, Michigan 48124
Email: amittal9@ford.com

## ABSTRACT

This research effort uses vehicular traffic flow techniques to model bicyclist longitudinal motion while accounting for bicycle interactions. Specifically, an existing car-following model, the Fadhloun-Rakha (FR) model is re-parametrized to model bicyclists. Initially, the study evaluates the performance of the proposed model formulation using experimental datasets collected from two ring-road bicycle experiments; one conducted in Germany in 2012, and the second in China in 2016. The validation of the model is achieved through investigating and comparing the proposed model outputs against those obtained from two state-of-the-art models, namely: the Necessary Deceleration Model (NDM), which is a model specifically designed to capture the longitudinal motion of bicyclists; and the Intelligent Driver Model, which is a car-following model that was demonstrated to be suitable for single-file bicycle traffic. Through a quantitative and qualitative evaluation, the proposed model formulation is demonstrated to produce modeling errors that are consistent with the other two models. While all three models generate trajectories that are consistent with empirically observed bicycle-following behavior, only the proposed model allows for an explicit and straightforward tuning of the bicyclist physical characteristics and the road environment. A sensitivity analysis, demonstrates the effect of varying the different model parameters on the produced trajectories, highlighting the robustness and generality of the proposed model.

# INTRODUCTION

As a combined result to an ever-increasing number of commuters and vehicles and the infeasibility of further capacity increases, congestion is becoming the most major problem facing modern urban cities worldwide. A popular solution adopted by policymakers to lessen traffic congestion in central downtown areas consists on advocating cycling as a sustainable commuting mode. That is justified by the fact that short-distance bike commuting often takes less time when accounting for congestion and delays in public transportation, and presents the most efficient way to increase the road capacity while maintaining existing infrastructure. For instance, several major cities are promoting the use of bikes between public transportation hubs and private transportation through the implementation of bike sharing systems. In fact, 119 US cities had a bike sharing system in 2017. Besides being a physical activity that is beneficial to human health, cycling has significant positive impacts on the environment as well. In fact, cycling results in significant reductions in fossil fuel consumption and vehicle emissions.

Despite the growing interest in bicycle use in the last decade and the urgent need to develop models and planning techniques for bicycle traffic operations, traffic researchers have minimally investigated the traffic flow dynamics of bicycles; unlike vehicular traffic flow, which is heavily studied. In fact, existing research that investigated bicycles as a means of transportation is relatively scarce [1-7]. The observed literature gap between vehicular and bicycle traffic research can be justified by the scarcity of naturalistic and experimental cycling data.

This research effort proposes to investigate the longitudinal motion dynamics of bicycles through the application of vehicular traffic techniques. The idea is based on the assumption that there are significant similarities between the traffic flow dynamics of bicycles and cars. The assumption is partly justified by the fact that existing cycling data comes from single-file ring-road experiments in which overtaking was not allowed. The team approach to achieve the study objective entails redesigning a car-following model to make it representative of a bike/bicyclist system rather than a vehicle/driver system. In that regard, two cycling datasets from German and Chinese experiments [7] will be used to re-parametrize the Fadhloun-Rakha (FR) car-following model [8] to make it representative of bicyclist acceleration/deceleration behavior. The reason behind the choice of the FR car-following model relates mostly to its ability to model the human-in-the-loop explicitly and separately from the vehicle dynamics. That aspect is anticipated to increase the chances of the resulting bike-following model in terms of capturing bicyclist variability, which is more influential than driver variability in car-following theory.

Concerning the paper layout, the paper will start by presenting the related work and literature including an overview of the Fadhloun-Rakha car-following model which will serve as the basis of the proposed model. Next, based on the assumption that significant similarities exist between car-following behavior and bicycle-following behavior, the proposed research will apply vehicular traffic flow modeling techniques to simulate bicyclist behavior, thus circumventing the challenges associated with modeling a very complex phenomenon from scratch. In that regard, the team will work on deriving a formulation for the proposed bike-following model from the FR car-following model [8]. That will be achieved through the re-parameterization of vehicle-related input variables along with the potential integration of necessary new parameters such that the characteristics and fundamentals of the bicycle/bicyclist system are fully captured. Thereafter, for validation purposes, the research team will assess the adequacy of the proposed formulation as a descriptor of bicycle longitudinal motion by estimating its quality of fit using two experimental datasets collected on a circular track in Germany [5] and in China [7]. The performance of the model will be assessed through comparing its performance against that of the Necessary Deceleration Model (NDM), which is a model specifically designed to capture the longitudinal motion of bicyclists; and the Intelligent Driver Model [1], which is a car-following model proved to be suitable for single-file bicycle traffic. That section also includes a sensitivity analysis that aims to illustrate the effect of varying the different model parameters on the produced trajectories. The analysis serves to highlight the robustness of the proposed model through exhibiting its ability to capture several characteristics related to both the bicyclist and the road environment, (e.g., gender, stamina, road grade). Finally, the conclusions of the papers are presented.

## BACKGROUND

Nowadays, the reliance of traffic engineering on computerized traffic simulations for planning, urbanization and environmental purposes, is increasing as a result of the continuous technological advancement and proliferation of microscopic simulation frameworks. While those computational tools allow the evaluation of different potential scenarios in a fast and cost-effective manner without dealing with any real world challenges, their results remain directly correlated to the accuracy and precision of the different logics integrated in them. That is the main reason for which a significant portion of traffic flow theory is oriented towards developing good descriptors of real traffic situations and empirical behavior. Looking at most of the existing microscopic simulation software, a main shortcoming that can be easily perceived relates to their orientation towards modeling vehicular traffic only. Such an observation becomes quite understandable when the huge gap between the number of research studies addressing vehicular and non-vehicular transportation modes is considered.

A main area of traffic flow theory for which the aforementioned gap is quite evident is car-following theory. Car-following theory proposes mathematical models [9-16] that aim to predict the temporal and spatial longitudinal behavior of a follower when the trajectory of the leader is known. In relation to modeling the longitudinal motion of moving entities, non-vehicular transportation modes such as cycling received little attention when compared to vehicular modes. The observed disparity is justified by the fact that the popularity of cycling as a sustainable commuting mode was minimal up until recently. Another potential reason relates to the lack of experimental and naturalistic data describing bicyclist behavior. In fact, it is only quite recently that datasets containing information about bicycle-following behavior became available to traffic researchers.

To illustrate the extent to which modeling the longitudinal motion of bicycles was ignored historically, there is no better argument than noting that the first model, specifically designed to simulate the following behavior of bicyclists, was only developed in 2012. The concerned model is the Necessary Deceleration Model (NDM) [5]. The NDM model is a discrete bicycle-following model that uses three components (*acc*, *dec₁*, *dec₂*) to compute the acceleration and the deceleration of a bicycle, as presented in Equations (1-5), using the following bicycle speed $v_n$, the spacing $s_n$, and the speed differential between the two bikes $\Delta v_n = v_n - v_{n-1}$.

$$a_{NDM} = acc - \min(dec_1 + dec_2, b_{max}) \qquad (1)$$

$$acc = \begin{cases} 0 & s_n \leq d(v_n) \\ \dfrac{v_f - v_n}{\tau} & s_n > d(v_n) \end{cases} \qquad (2)$$

$$dec_1 = \begin{cases} min\left(\dfrac{(\Delta v_n)^2}{2(s_n - s_j)}, b_{max}\right) & \Delta v_n < 0 \\ 0 & \Delta v_n \geq 0 \end{cases} \quad (3)$$

$$dec_2 = \begin{cases} b_{max}\left(\dfrac{s_n - d(v_n)}{l_n - d(v_n)}\right)^2 & s_n \leq d(v_n), \Delta v_n \geq -\varepsilon \\ 0 & otherwise \end{cases} \quad (4)$$

$$d(v_n) = s_j + T \cdot v_n \quad (5)$$

Besides the free-flow speed $v_f$ and the spacing at jam density $s_j$, the NDM requires the calibration of three additional parameters, which are: a constant of proportionality $T$, a maximum deceleration level $b_{max}$, and a relaxation time $\tau$ that controls how fast a bicycle accelerates to the desired speed $v_f$. Furthermore, the model involves the use of the bicycle length $l_n$ which is set equal to 1.73 m (average bicycle length) and a positive constant $\varepsilon = 0.5$ m/s.

Based on the observation that there are no major differences between the dynamics of single-file bicycle traffic and vehicular traffic, another approach used by researchers to model the longitudinal motion of bicycles consisted on investigating the possibility of capturing cyclists' behavior through revamping certain aspects of existing car-following models. That is the case of the Intelligent Driver Model (IDM) [16] which, after a simple re-parameterization of its parameters, was proven to be a good descriptor of bicycle-following behavior [1]. In a similar fashion to the NDM model, the IDM model is formulated as a set of coupled ordinary differential equations as presented in Equations (6-7) and requires the calibration of five parameters.

$$a_{IDM}(v_n, s_n, \Delta v_n) = a\left(1 - \left(\dfrac{v_n}{v_f}\right)^4 - \left(\dfrac{s^*(v_n, \Delta v_n)}{s_n}\right)^2\right) \quad (6)$$

With

$$s^*(v_n, \Delta v_n) = s_j + v_n T + \dfrac{v_n \Delta v_n}{2\sqrt{a.b}} \quad (7)$$

Where $s^*$ denotes the steady state spacing, $a$ is the maximum acceleration level, $b$ is the maximum deceleration level, and $T$ is the desired time headway.

## METHODOLOGY AND FORMULATION

This section starts with an overview of the FR car-following model that is followed by a description of the methodologies leading to the development of the redesigned bicycle-following model formulation.

### Fadhloun-Rakha Model

One of the simplest car-following strategies entails attempting to follow the lead vehicle at a constant headway, which is typically taken equal to the driver perception-reaction time $T$, as illustrated in Equation 8. This model is also known as the Pipes or GM-1 model [17-19]. This time headway ensures that the subject vehicle $n$ follows its leader at a safe spacing in order to avoid a collision under steady-state conditions (i.e. when both vehicles are traveling at the same constant velocity and assuming that the subject vehicle's deceleration maneuver starts $T$ seconds after the lead vehicle decelerates).

$$\tilde{s}_n = s_j + T v_n \quad (8)$$

In the context of car-following modeling, Van Aerde [20] and Van Aerde and Rakha [21] proposed a more general formulation that reflects empirical driver behavior better than other models. This formulation combines the Pipes (Equation 8) and the Greenshields models to generate a more general formulation [22-24], presented in Equation 9.

$$\tilde{s}_n = c_1 + \frac{c_2}{(v_f - v_n)} + c_3 v_n \qquad (9)$$

Here, $c_1, c_2,$ and $c_3$ are model coefficients that can be computed using key roadway traffic stream parameters (Equation 10) [22], namely: the free-flow speed, $v_f$; the speed-at-capacity, $v_c$; the roadway capacity, $q_c$; and the roadway jam density, $k_j$ (the inverse of the jam density spacing, $s_j$).

$$c_1 = \frac{v_f}{k_j v_c^2}(2v_c - v_f);$$
$$c_2 = \frac{v_f}{k_j v_c^2}(v_f - v_c)^2; \qquad (10)$$
$$c_3 = \frac{1}{q_c} - \frac{v_f}{k_j v_c^2}$$

If the lead vehicle is traveling at a lower velocity than the following vehicle (non-steady-state conditions) then the desired safe following spacing can be computed using Equation 11. This allows the following driver to drive at a spacing longer than the steady-state spacing when the vehicle ahead of it is driving at a lower speed.

$$\tilde{s}_n = \max\left(c_1 + \frac{c_2}{(v_f - v_n)} + c_3 v_n + \frac{v_n^2 - v_{n-1}^2 + \sqrt{(v_n^2 - v_{n-1}^2)^2}}{4d}, s_j\right) \qquad (11)$$

Finally, the vehicle acceleration behavior is governed by the vehicle dynamics, as demonstrated in Equation 12 to ensure that vehicle accelerations are realistic.

$$a_{max} = \frac{\min\left(\frac{\beta \eta_d P_n}{v_n}, m_{ta} g \mu\right) - \frac{\rho C_d C_h A_f v_n^2}{2} - m g C_{r0}(C_{r1} v_n + C_{r2}) - m g G}{m} \qquad (12)$$

Rakha and Lucic [25] introduced the $\beta$ factor in order to account for the gearshift impacts at low traveling speeds when trucks are accelerating. This factor is set to 1.0 for light-duty vehicles [26]. Other parameter definitions are: $\eta_d$ is the driveline efficiency (unitless); $P$ is the vehicle power (W); $m_{ta}$ is the mass of the vehicle on the tractive axle (kg); $g$ is the gravitational acceleration (9.8067 m/s²); $\mu$ is the coefficient of road adhesion or the coefficient of friction (unitless); $\rho$ is the air density at sea level and a temperature of 15°C (1.2256 kg/m³); $C_d$ is the vehicle drag coefficient (unitless), typically 0.30; $C_h$ is the altitude correction factor equal to 1-0.000085$h$, where $h$ is the altitude in meters (unitless); $A_f$ is the vehicle frontal area (m²), typically 0.85 multiplied by the height and width of the vehicle; $C_{r0}$ is a rolling resistance constant that varies as a function of the pavement type and condition (unitless); $C_{r1}$ is the second rolling resistance constant (h/km); $C_{r2}$ is the third rolling resistance constant (unitless); $m$ is the total vehicle mass (kg); and $G$ is the roadway grade (unitless).

To capture the driver input the $f_p$ factor is introduced, which ranges between 0.0 and 1.0. The final FR model formulation considering the deceleration to avoid a collision with a slower traveling leader is cast using Equation 13. This equation includes two terms. The first term is the acceleration term while the second term is the deceleration term. Both terms ensure that the following vehicle does not collide with its leader.

$$a_n = f_p a_{max} - \frac{\left[v_n^2 - v_{n-1}^2 + \sqrt{(v_n^2 - v_{n-1}^2)^2}\right]^2}{16(d_{des} - gG)(s_n - s_j)^2} \tag{13}$$

Here $f_p$ is computed using Equation 14 where $X_n$ is calculated using Equation 15.

$$f_p = e^{-g_1 X_n}\left(1 - X_n^{g_2} e^{g_2(1 - X_n)}\right)^{g_3} \tag{14}$$

$$X_n = \frac{\min(\tilde{s}_n, \tilde{s}_n((1-\alpha)v_f))}{\min(s_n, \tilde{s}_n((1-\alpha)v_f))} \cdot \frac{v_n}{\tilde{v}_n} \tag{15}$$

Here $\tilde{s}_n$ is the desired spacing for the current speed (computed using Equation 11); $\tilde{v}_n$ is the desired speed for the current spacing (which is computed by solving for the driver's desired speed based on its current spacing using Equation 11); $\alpha$ is the percentile off $v_f$ (suggested to be 2.5%); $d_{des}$ is the desired deceleration level; $g_1$, $g_2$ and $g_3$ are model parameters that are calibrated to a specific driver, and model the driver power input through the application of the gas pedal.

In order to ensure that the parameters ($g_1$, $g_2$ and $g_3$) result in a minimal maximum value of $f_p$ in the deceleration domain, an iterative procedure was developed. The iterative procedure, presented in Equation 16, is only approximate and converges relatively fast (within four to five iterations) to the location of the maximum of $f_p$, which is then verified to be below a threshold $\varepsilon$ (for instance, $\varepsilon = 0.1$). By doing so, it is decided whether the chosen values for the $g_1$, $g_2$ and $g_3$ parameters are accepted or rejected. Of course, this procedure was only adopted after ensuring that the number of the different combinations of ($g_1$, $g_2$ and $g_3$) that would result in $f_p(X_{k \to inf}) < \varepsilon$ is significant.

$$\begin{cases} X_0 = 3\left(-1 + \sqrt{2\ln(3)}\right) \\ X_{k+1} = 3\left[-1 + \sqrt{2\ln\left(3\left[1 + \frac{g_2 g_3}{g_1}\left(1 - \frac{1}{X_k}\right)\right]^{1/g_2}\right)}\right] \end{cases} \tag{16}$$

Finally, three noise variables are added to the model formulation in order to capture the perception and control inaccuracies of the drivers. The first two signals attempt to model the perception errors in estimating the leader's speed and the gap distance separating the two vehicles. They consist of two Wiener processes that are incorporated in the model formulation as presented in Equations 17 and 18. On the one hand, Equation 17 emulates the driver's inability to have an exact estimation of the speed of the leading vehicle. On the other hand, Equation 18 simulates the error committed while estimating the spacing separating them. Additionally, a white noise signal, presented in Equation 19, is added to the model's expression to capture the control errors during the acceleration and deceleration maneuvers. The compounding effect of these three signals makes the model output more representative of human behavior. The model output is computed as the sum of Equation 19 and Equation 13 in which $\widetilde{u_n}(t)$ and $\widetilde{s_{n+1}}(t)$ are used instead of $u_n$ and $s_{n+1}$.

$$\begin{cases} \widetilde{u_{n-1}}(t) = u_{n-1}(t - \Delta t) - 0.01(s_n - s_j)\left(e^{-0.01} \cdot W_l(t - \Delta t) + \sqrt{0.02} \cdot N(0,1)\right) \\ W_l(1) = N(0,1) \end{cases} \tag{17}$$

$$\begin{cases} \breve{s}_n(t) = s_n(t - \Delta t) \times e^{0.1\left(e^{-0.01 \cdot W_s(t-\Delta t)} + \sqrt{0.02} \cdot N(0,1)\right)} \\ W_s(1) = N(0,1) \end{cases} \quad (18)$$

$$\breve{a}_n(t) = N(0, 0.25) \quad (19)$$

**Bicycle Model Formulation**
For the most part, the car-following strategy of the FR model remains valid for modeling the longitudinal single-file motion of bicycles. Specifically, the functions governing collision avoidance, steady state behavior, and human behavior modeling would have the same functional forms. For the aforementioned functions, the differences between vehicular traffic and bicycle traffic would be expressed at the level of the adopted values of their different parameters. The latter is not the case for the vehicle dynamics model [27] which requires the implementation of structural modifications in order to make it descriptive of the maximum acceleration behavior of bicycles.

The biggest challenge that faced the research team in this phase related mostly to choosing an adequate expression for the tractive force. Having a good approximation of the traction as a result of pedaling, significantly impacts the precision and the accuracy of the bicycle trajectories generated by the model as it defines the maximum acceleration profile $a_{max}$. The proposed expression for the tractive force was achieved by modeling the cyclist as a motor delivering power. In order to account for the cyclist output variability over time, we opted to estimate the power as the product of the cyclist weight and the highest average power that can be sustained over a certain period of time, commonly known as the functional threshold power (FTP factor in W/kg). Understandably, the FTP factor depends on several variables such as gender, stamina, and the time interval as shown in TABLE 1. For instance, a male cyclist in a good shape is able to generate an average of 3.91 W/kg over an hour period and a higher average of 8.28 W/kg over a 5-minute period. For a female cyclist in the same shape, these values decrease slightly to 3.39 W/kg and 6.75 W/kg over the same time periods. Finally, in order to account for the losses incurred while the pedaling power is transmitted to the rear wheel, several efficiency factors are applied. These factors attempt to model the effect of the bicycle gears and the friction at the level of the bicycle chain. TABLE 2 presents a summary of the needed changes to account for the differences between bicycles and vehicles.

**TABLE 1 Maximal power outputs for different cyclist categories [28]**

| Bicyclist Condition | Male | | | Female | | |
|---|---|---|---|---|---|---|
| | 1 min | 5 min | 1 hour | 1 min | 5 min | 1 hour |
| **World Class** | 11.50 | 7.60 | 6.40 | 9.29 | 6.61 | 5.69 |
| **Exceptional** | 10.35 | 6.57 | 5.51 | 8.38 | 5.68 | 4.87 |
| **Excellent** | 9.66 | 5.95 | 4.98 | 7.84 | 5.13 | 4.38 |
| **Very good** | 8.97 | 5.33 | 4.44 | 7.3 | 4.57 | 3.88 |
| **Good** | 8.28 | 4.70 | 3.91 | 6.75 | 4.02 | 3.39 |
| **Moderate** | 7.48 | 3.98 | 3.29 | 6.12 | 3.37 | 2.82 |
| **Fair** | 6.79 | 3.36 | 2.75 | 5.57 | 2.82 | 2.32 |
| **Untrained** | 5.87 | 2.53 | 2.04 | 4.85 | 2.07 | 1.67 |

**TABLE 2 Summary of the dynamics model for bicycles and cars**

| | Bicycles | Cars |
|---|---|---|
| **Tractive Force** | $\min\left(\eta_{eff}\eta_{gears}\frac{m_{cyclist}P_{ftp}}{v_n}, m_{ta}g\mu\right)$<br>$\eta_{eff}$ depends on bicycle chain<br>$\eta_{gears}$ depends on gears and bike geometry<br>$m_{ta}$ depends on center of gravity position | $\min\left(\frac{\beta\eta_d P_n}{v_n}, m_{ta}g\mu\right)$ |
| **Rolling Resistance** | $mgC_{rr}$<br>$C_{rr}$ depends on road type | $mgC_{r0}(C_{r1}v_n + C_{r2})$ |
| **Aerodynamic Resistance** | Same formulation except:<br>$C_dA_f$ depends on cyclist physique and posture on bike | $C_dA_f$ depends on car shape |
| **Grade Resistance** | Same formulation | |

## ANALYSIS

In this section, we propose to evaluate the performance of the proposed model formulation against the NDM and the IDM models using the German and Chinese datasets. By doing so, the suitability of the reparametrized FR model for simulating bicycle traffic flow and following behavior is investigated.

**Experimental Data**
The bicycle trajectory data used in this study is the result of two controlled ring-road experiments. The first experiment was carried out by the University of Wuppertal in conjunction with Julich Supercomputing Center in May, 2012 [5]. Participants from different age groups were instructed to follow one another without overtaking on an 86-meter circular track. The experiments were conducted with 5, 7, 10, 18, 22 and 33 cyclists allowing them to capture the effects of different density levels. Through the use of two cameras overlooking the road, the cyclists' trajectories were captured along a 20 m long straight section of the test track. As a side remark, it is worth mentioning that the experiments were mainly performed for the sake of calibration and validation of the NDM model, which is the first model designed to simulate the longitudinal motion of bicycles.

The second dataset originates from a 2016 Chinese experiment that was conducted on a 146-meter circular track. Like the German experiments, the experimental runs were performed with 39, 48 and 63 cyclists for the purpose of reproducing different global densities. Using a video camera that was mounted on the top of a high building, the researchers were able to extract the full trajectories of the bicycles.

The aggregated data for the different runs can be visualized in FIGURE 1 (German experiment) and FIGURE 2 (Chinese experiment), which shows the data in different domains of the fundamental diagram. The dispersion of the data confirms a similar behavior to that originating from vehicular traffic. The macroscopic data points presented in the figure are obtained from the microscopic bicycle trajectories through the use of virtual loop detectors located at the entrance and the exit of the 20 m section on which the different trips occurred for the German experiment, and on a 50 m section of the circular track for the Chinese experiment.

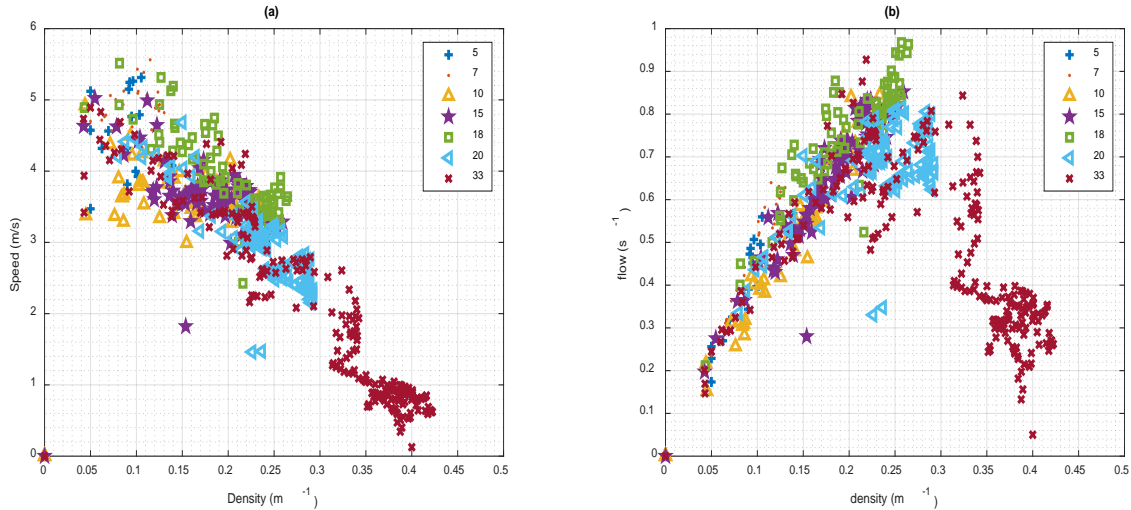

**FIGURE 1 German experiment a) Speed-density relationship; b) Flow-density relationship**

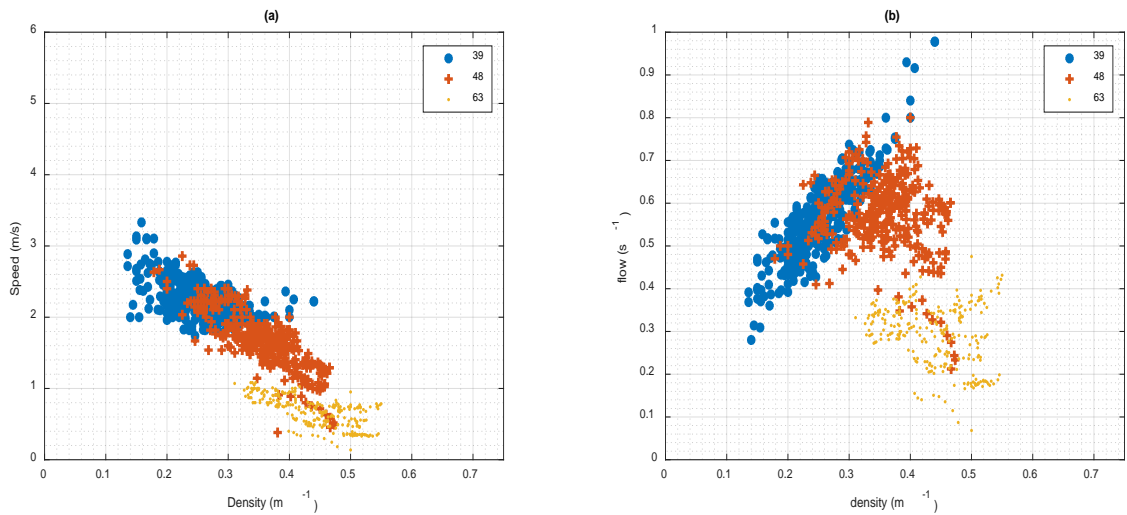

**FIGURE 2 Chinese experiment a) Speed-density relationship; b) Flow-density relationship**

## Calibration

For each of the studied models, a number of inputs are needed. These inputs can be categorized into three groups. The first category includes the inputs that are extracted directly from the datasets such as the time-space and the time-speed profiles of the leading vehicle, the starting location and speed of the following vehicle. The second category comprises the parameters that are calibrated using all the trips as they are more representative of the road facility and the fundamental diagram that governs it rather than the trip itself. Subsequently, these parameters are assigned a single value across all of the bicycle-following events in order to maintain and represent the homogeneity of the road facility. Namely, the concerned parameters are the free-flow speed $u_f$ and the spacing at jam density $s_j$ (inverse of $k_j$), which are shared among the three models, along with the roadway capacity $q_c$, and the speed-at-capacity $u_c$, which are needed to generate a simulated trajectory in the case of the proposed model formulation. Those parameters were estimated using the calibration procedure proposed by Rakha and Arafeh [29] using the bulk macroscopic data shown in FIGURE 1 and FIGURE 2. The reasons for which the data related to the two experiments was combined are twofold: First, the Chinese experiment shows a clear lack of data in the free-flow regime. Second, the two datasets seem to overlap almost perfectly as shown in FIGURE 3, which presents the calibration results of the fundamental diagram of the facility.

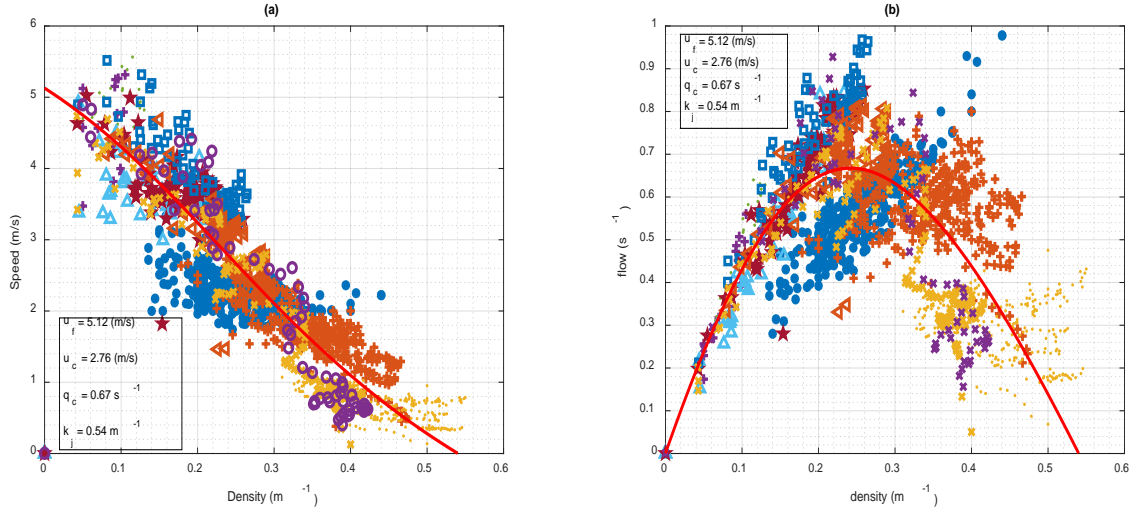

**FIGURE 3 Calibration of the fundamental diagram a) Speed-density relationship; b) Flow-density relationship**

Next, the calibration procedure is complemented by the calibration of the remaining variables. At a first glance, the calibration of the proposed model might seem complex due to the sheer number of variables involved. However, most of these parameters have fixed values that are either constant or dependent on certain characteristics of the road and/or the bicycle. For instance, knowing that the experiments were run on a dry and flat asphalt road, the values of the grade $G$, the rolling coefficient $C_{rr}$, and the friction coefficient $\mu$ were set to 0, 0.004, and 0.8 respectively. For the remaining variables, the following assumptions are made:

- The cyclist is a male in a good shape ($P_{ftp}$=3.91 W/kg),
- The bicycle weighs 8 kg,
- The proportion of the total mass on the rear axle equals 0.60,
- The aerodynamics coefficients are such that $C_d A_f = 0.4$,
- The desired deceleration level $d_{des}$ is equal to 1.5 m/s²,
- The total efficiency factor $\eta_{eff}\eta_{gears} = 0.62$.

To highlight the practicality and ease of implementation of the model, we present in Equation 20, a simplified formulation of the bicycle dynamics model in which the different parameters were substituted with their respective values. Unless a potential user is interested in modeling a very specific scenario with a high level of detail, we judge that the presented version of the dynamics model is adequate for testing purposes of the model for the average user once implemented in Equation 13.

$$a_{max} = \min\left(2.42 \frac{m_{cyclist}}{(m_{cyclist} + 8).v_n}, 4.7\right) - \frac{0.003v_n^2 + 0.04}{(m_{cyclist} + 8)} \quad (20)$$

Considering the above along with a cyclist weight of 75 kg, each of the considered models would require the calibration of three additional parameters. In that context, the values for ($g_1$, $g_2$, $g_3$) for the proposed model, ($\tau, T, b_{max}$) for the NDM model and ($a, T, b$) for the IDM model are obtained, for each cyclist trajectory, through an optimization operation that aims to minimize the error between empirical and simulated data. For each pair of successive trajectories (leader/follower), the simulated trajectory of the follower was initialized with the empirical speed and spacing for each of the three models. Next, the calibration procedure of each model was conducted heuristically with the objective of finding the set of parameters resulting in the smallest error values. In that context, the absolute spacing error, presented in Equation 21, was chosen to serve as the error objective function given that this was one of the errors used to compare the NDM and IDM models in a previous publication [1]. Subsequently, the performance of each model is quantified through the comparison of the simulated spacing data $s^{sim}$ against the empirical observations $s^{obs}$.

$$f_{abs} = \frac{\sum_{i=1}^{n}\left(s_i^{sim} - s_i^{obs}\right)^2}{\sum_{i=1}^{n}\left(s_i^{obs}\right)^2} \tag{21}$$

The choice to optimize each model with regards to the absolute spacing error is judged reasonable given that the optimization operation was done on an event-by-event basis. We opted to calibrate each model separately for each event rather than for the dataset as a whole. Even though that increased the computation time, a more fair comparison of the results is made possible as each model was allowed to propose its best possible fit for each trajectory. Hence, the different model outputs are reflective of the strength points of each model. The use of the denominator in the error objective function is justified by the need to account for the trip duration variability across the different events. By dividing the absolute error with a variable that is sensitive to the aforementioned variability; its effects on the chosen error metric are minimized.

Furthermore, given the relatively long duration of the events specific to the Chinese experiment, we have opted to calibrate the three models using the initial 75% of each trajectory for calibration purposes. The remaining portion of the trajectory (25%) is used to validate the models and quantify their predictive power. Such a procedure was not possible for the German data, for which the integrality of the trajectory is used, due to the very short trips involved.

**Results**
Having access to the calibrated parameters, the simulated trajectories were obtained for the different events of the two datasets. The characteristics of the calibration errors related to the German and Chinese experiments are presented in TABLE 3. For the German data, lower error values are observed for the IDM and the NDM models when compared to the FR model. However, the trend is reversed when the Chinese data is considered. In fact, the FR and NDM models are shown to outperform the IDM model with the FR model results slightly better than those of the NDM. The observed discrepancies between the results of the two datasets could be contributed to the short trip durations in the German experiments in which the trajectories are collected only over a 20 meter section (in comparison to the 146 meter trajectories of the Chinese experiment).

**TABLE 3 Characteristics of the calibration errors for the FR, IDM and NDM models**

|         | German Dataset |        |        | Chinese Dataset |       |       |
|---------|----------------|--------|--------|-----------------|-------|-------|
|         | FR             | IDM    | NDM    | FR              | IDM   | NDM   |
| Mean    | 0.0063         | 0.0025 | 0.0033 | 0.043           | 0.058 | 0.044 |
| Median  | 0.0024         | 0.0004 | 0.0005 | 0.029           | 0.040 | 0.030 |
| Std Dev | 0.0096         | 0.0077 | 0.0100 | 0.064           | 0.076 | 0.052 |

   Further error metrics for the IDM, NDM and FR models are shown in TABLE 4, which presents the key distribution parameters for the error objective function of the Chinese trajectory data. For instance, when the full trajectories are considered, the table demonstrates that the proposed model along with the NDM model offer almost identical percentiles and that they are slightly better in terms of fitting the observed data than the IDM model. That statement is further supported by FIGURE 4, which plots the empirical cumulative distribution functions corresponding to the three models. Specifically, FIGURE 4.c demonstrates that the cumulative distribution functions of the FR model and the NDM models are almost identical at every data point of the error axis. Next, the distribution functions corresponding to the German experimental data are plotted in FIGURE 5. While the plot makes it clear that the FR model underperforms the other two models, the small magnitude of the errors involved does not undermine the performance of the model or its suitability for modeling bicycle following behavior.

**TABLE 4 Distribution characteristics of the error function**

|                     |     | 25% percentile | Median | 75% percentile | 95% percentile |
|---------------------|-----|----------------|--------|----------------|----------------|
| **Calibration**     | FR  | 0.017          | 0.029  | 0.047          | 0.125          |
|                     | IDM | 0.023          | 0.040  | 0.067          | 0.172          |
|                     | NDM | 0.017          | 0.030  | 0.053          | 0.130          |
| **Validation**      | FR  | 0.028          | 0.054  | 0.118          | 0.396          |
|                     | IDM | 0.026          | 0.054  | 0.120          | 0.505          |
|                     | NDM | 0.025          | 0.048  | 0.108          | 0.463          |
| **Full Trajectory** | FR  | 0.023          | 0.036  | 0.062          | 0.158          |
|                     | IDM | 0.027          | 0.044  | 0.076          | 0.194          |
|                     | NDM | 0.022          | 0.036  | 0.063          | 0.154          |

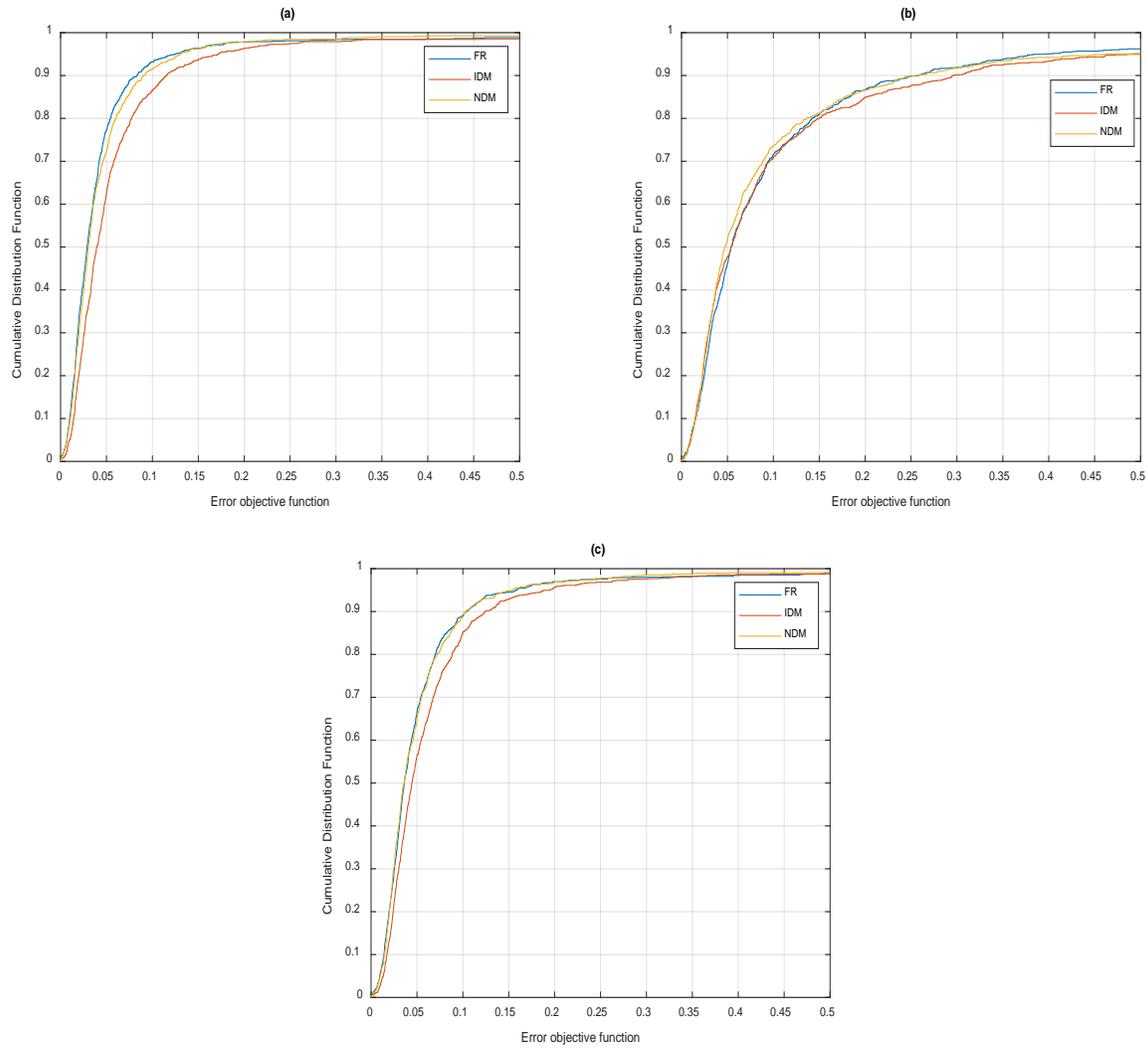

**FIGURE 4 Cumulative distribution function of the error of the Chinese dataset using: a) the first 75% of the trajectory (calibration trajectory); b) the last 25% of the trajectory (validation trajectory); c) the full trajectory**

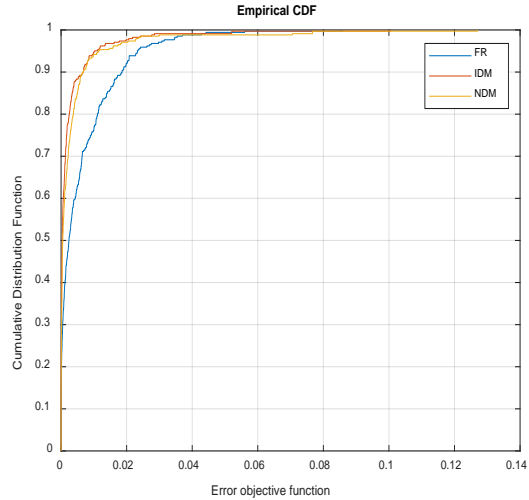

**FIGURE 5 Cumulative distribution function of the error of the German dataset**

To better statistically quantify the difference in performance between the different models, the rank of each model was determined for each event based on the calculated error objective function. In doing so, it was found that the FR model ranked first as it outperformed the NDM and the IDM models in approximately 47% of the cases. The NDM model demonstrated a good performance as well as it offered the best fit in 42% of the events. To have a deeper understanding of the results, the performance of the FR model was compared face-to-face with the NDM model. It was found that the two models performed equally well with each model producing the lowest error in exactly half the events. That is in accordance with the results presented in TABLE 4 and FIGURE 4.

While all the models resulted in a good fit to the data without any clear outperformer, it is noteworthy to mention the main advantage of the proposed model. Specifically, the main advantage of the FR model lies in its ease of tuning to capture different dynamic characteristics specific to both the cyclist and the environment such as athletic capability, size, gender, weather, and road grade. The robustness of the model is further complemented by its explicit inclusion of parameters that are reflective of the human-in-the-loop element separately from the bicycle dynamics variables. Put simply, the model is able to emulate cyclist behavior variability even when the same physical characteristics and road conditions are considered. To illustrate the previous points, the study proceeded to perform a simple sensitivity analysis in which a 100-meter trip is simulated for different case scenarios. In each of the scenarios, all the model parameters were set to a fixed value except for one. That would allow visualizing its impact on the generated trajectories. The following case scenarios are considered:

- *Scenario 1*: the coefficient of friction of the road, $\mu$ is varied between 0.1 and 0.8 in order to model several road conditions ranging from icy to dry as shown in FIGURE 6.a.
- *Scenario 2*: the functional threshold power is varied between 2.0 W/kg to 6.4 W/kg in order to model the physical capability of the eight cyclist categories ranging from an untrained individual to a world-class athlete (FIGURE 6.b).
- *Scenario 3*: several road grades were investigated ranging from a 4% downhill to a 4% uphill at 2% increments (FIGURE 6.c).

- *Scenario 4*: The final scenario investigates the effect of the cyclist gender on the results (FIGURE 6.d).

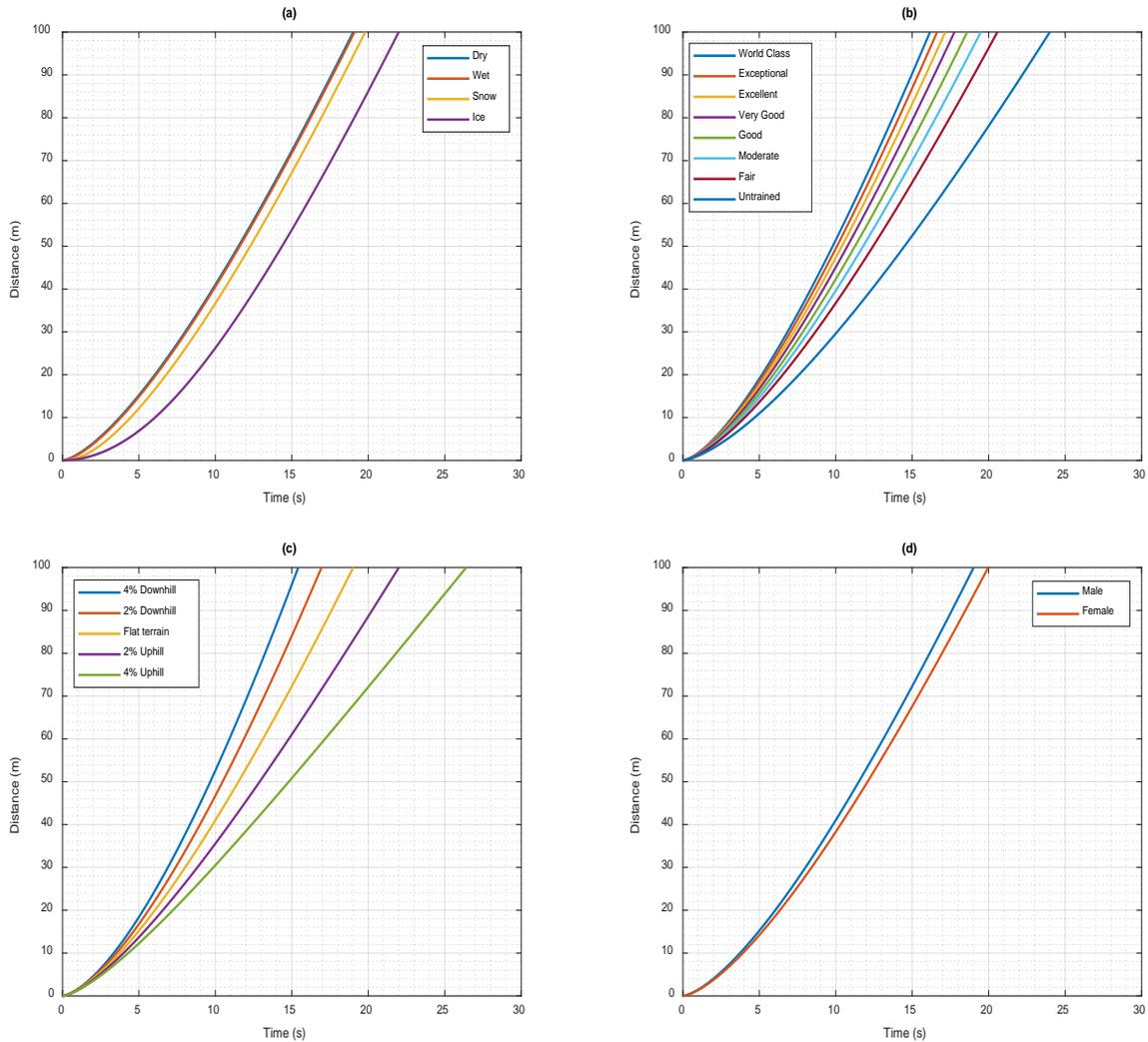

**FIGURE 6 Effect of different model parameters on the simulated trajectories: a) Coefficient of road friction; b) Functional threshold power; c) Road Grade; d) Gender**

Back to the calibration results, the possibility of the existence of any relationships and/or correlation between the different combinations of optimal parameters ($g_1$, $g_2$, $g_3$) was explored. FIGURE 7.a to FIGURE 7.c, which plots the variation of $g_1$, $g_2$ and $g_3$ against each other, makes it easy to identify certain patterns related to the range of variation of those parameters responsible for modeling cyclist behavior variability. First, it is relatively clear in FIGURE 7.a and FIGURE 7.c that there is a significant concentration of $g_2$ values in the area between 0 and 100. Second, the plots suggest that $g_1$ is the easiest variable in terms of calibration as the range of the optimized values is quite limited compared to the other two variables. Another observation relates to a potential correlation between the $g_2$ and $g_3$ parameters. The observed patterns in FIGURE 7.c seem to not be random and suggest the existence of a family of hyperbolic functions that governs the relationship between the two parameters. Determining the latter functions and analytically confirming the observed boundaries would probably result in the reduction of the computational time of the optimal solution in comparable future studies. An analytical investigation of the relationships between the three parameters would constitute an interesting and useful complement to this study.

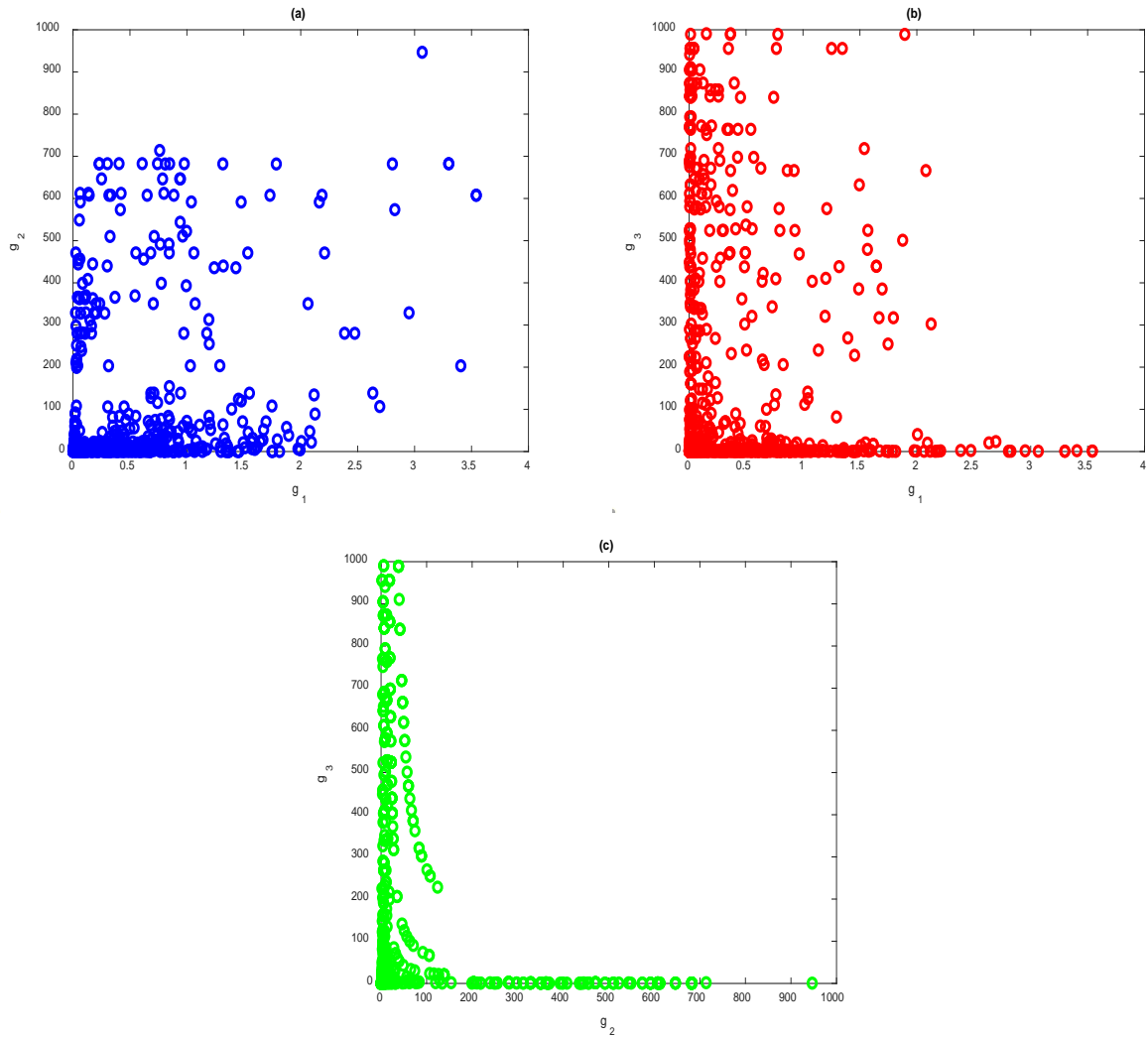

**FIGURE 7 Variation of the FR model parameters against each other: a) g1 vs. g2; b) g1 vs. g3; c) g2 vs. g3**

From a qualitative standpoint, FIGURE 8 presents plots of both the empirical and the simulated time-space trajectories corresponding to the Chinese experimental run in which 63 cyclists were involved. The choice of that specific scenario is not random and is justified by the presence of stop-and-go waves and several traffic perturbations due to the high density level. That would allow for a more comprehensive evaluation of the suitability of the proposed FR model formulation for the simulation of bicycle-following behavior. The figure confirms that the FR model reflects the overall behavior of the cyclists. Specifically, the simulated trajectories are shown to successfully follow the different patterns in the empirical data, and to precisely replicate the empirically observed stop-and-go waves.

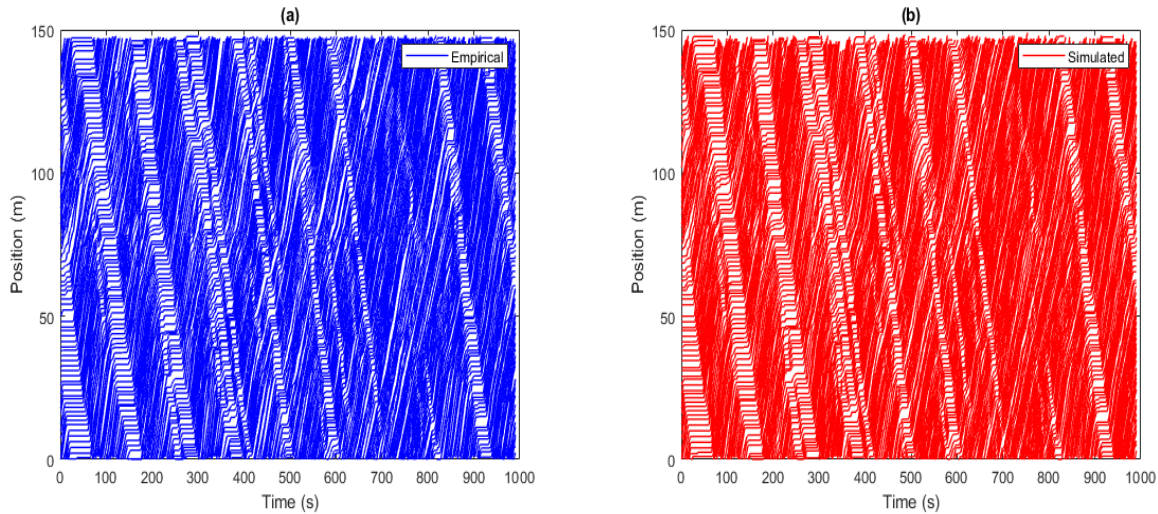

**FIGURE 8 Bicycle trajectories for the Chinese experimental run with 63 cyclists: a) Empirical; b) Simulated**

## CONCLUSIONS

This paper extended the FR car-following model making it suitable for the simulation of the longitudinal motion of bicycles. The extension was achieved through the re-parameterization of vehicle-related input variables along with the integration of new parameters such that the characteristics and fundamentals of the bicycle/bicyclist system are fully captured. The proposed model is the first point-mass dynamics-based model for the description of the following behavior of bicycles in both constrained and unconstrained conditions. The main benefit of the model lies in its robustness and its ability to model bicyclist behavior variability. The proposed model is the only existing model that is sensitive to the bicyclist physical characteristics and the roadway surface conditions.

The study used experimental ring-road data to validate the proposed model through comparing its performance against two other state-of-the-art bicycle following models. The research findings demonstrate that the FR model is successful in replicating empirical bicyclist behavior and that the techniques used in car-following theory could be used, with minor modifications, to model bicyclist longitudinal motion modeling; thus eliminating the need to develop bicycle-specific models.

## ACKNOWLEDGMENTS

The authors acknowledge the financial support provided by the University Mobility and Equity Center (UMEC) and funding from the Ford Motor Company.

## AUTHOR CONTRIBUTIONS

The authors confirm contribution to the paper as follows: study conception and design: H.A. Rakha and K. Fadhloun; data collection: K. Fadhloun; analysis and interpretation of results: K. Fadhloun and H.A. Rakha; draft manuscript preparation: K. Fadhloun, H. Rakha, and A. Mittal. All authors reviewed the results and approved the final version of the manuscript.

Fadhloun, Rakha and Mittal                                                                                                                    18